\documentclass[allclo]{FBSart}
\usepackage{amsfonts}
\usepackage{amssymb}
\usepackage[dvips]{graphicx}

\title{Bright vortex solitons in Bose Condensates}
\author{Sadhan K. Adhikari\thanks{\textit{E-mail address:} 
adhikari@ift.unesp.br}}
\institute{Instituto de F\'isica Te\'orica, Universidade Estadual
Paulista, \\ 01405-900 S\~ao Paulo, S\~ao Paulo, Brazil}

\runningauthor{S. K. Adhikari}
\runningtitle{Bright vortex solitons in Bose-Einstein Condensates}
\begin{document}

\maketitle
\begin{abstract}
We suggest  the possibility of observing and studying  bright
vortex solitons
in attractive Bose-Einstein condensates in three dimensions with a
radial trap. Such systems lie on the verge of
critical stability and we discuss the conditions of their stability. 
We study the interaction between two such solitons. Unlike the
text-book solitons in one dimension, the interaction between two radially
trapped and axially free three-dimensional solitons is inelastic in nature
and involves exchange of particles and deformation in shape.
The interaction remains repulsive for all phase $\delta$
between them except for $\delta\approx 0$.
\end{abstract}

\section{Introduction}

Solitary waves or solitons are a consequence of nonlinear dynamics. A
classic text-book example of soliton appears  in the  following
one-dimensional
nonlinear free Schr\"odinger equation in
dimensionless units
\begin{equation} \label{1}
\left[-i
 \frac{\partial }{\partial t}
-  \frac    {\partial^2 }{\partial y^2} -
| \Psi(y,t)|^2 \right]
\Psi(y,t)  =0.
\end{equation}
The solitons of this equation  are localized solution due to the
attractive
nonlinear
interaction $-| \Psi(y,t)|^2 $
with wave function  at time
$t$ and position
$y$:  $\Psi(y,t)= \sqrt{2|\Omega| }\exp(-i\Omega t){\mbox{sech}} (y\sqrt
{|\Omega|})$, with $\Omega$ the energy \cite{7}.  

Solitons have been noted in  optics \cite{0},
high-energy physics and water waves \cite{1}, and more recently in
Bose-Einstein
condensates (BEC) \cite{3,4}. 
The Schr\"odinger equation
with a nonlinear interaction $-|\Psi|^2$ does not sustain a localized
solitonic solution in three dimensions. However, a radially trapped and
axially free version
of this equation in three dimensions does sustain such a bright solitonic
solution \cite{4a0} which has been observed experimentally  
\cite{3,4}. Here we study the
dynamics of these bright solitons. We also suggest  that such solitons can
be generated in an  axially rotating nonzero angular momentum state, which
are called bright vortex solitons and can be observed in BEC. 

A   number of bright solitons constituting a
soliton train was
  observed in an experiment  by Strecker {\it et al.}
\cite{3}, where they turned a repulsive BEC of $^7$Li
atoms attractive  by manipulating
the background magnetic field near a    Feshbach
resonance \cite{4a}. It was found
\cite{3} that solitons
in such a train usually stay apart. Also, often a soliton was found
to be missing from a train \cite{3}. 
There have been theoretical attempts
\cite{6,6a0,6a01} to simulate essentials of these experiments
\cite{3,4}.

We use the explicit numerical solution of the
axially-symmetric  mean-field Gross-Pitaevskii (GP) equation \cite{8} to
study
the dynamics of bright solitons in a soliton train
\cite{3}. 
Attractive BEC's may not form vortices in a thermodynamically stable
state.  However, due to the conservation of angular momentum, a vortex
soliton
train could be generated by suddenly changing the inter-atomic interaction
in an axially-symmetric rotating vortex condensate from repulsive to
attractive near a Feshbach resonance \cite{4a} in the same fashion as in
the
experiment by Strecker {\it et al.} \cite{3} for a non-rotating BEC.
Alternatively, a single vortex soliton could be prepared and studied in
the laboratory by forming a vortex in a small repulsive condensate and
then making the interaction attractive via a Feshbach resonance and
subsequently reducing the axial trap slowly.

\section{Mean-field Model and Results}

{\underline{Mean-field Model:}} In a quantized vortex state \cite{9}, with
each atom having angular
momentum $L\hbar$ along the axial $y$ axis, the axially-symmetric  wave
function
can be
written as  $\Psi({\bf r}, \tau)=
\varphi(r,y,\tau)\exp (iL\theta) $ where $\theta$ is the azimuthal
angle and $r$ the radial direction. The dynamics of the BEC in an
axially-symmetric trap can be
described by the following GP equation \cite{8,9}
\begin{eqnarray}\label{d1} &\biggr[&-i\frac{\partial
}{\partial t} -\frac{\partial^2}{\partial
r^2}+\frac{1}{r}\frac{\partial}{\partial r} -\frac{\partial^2}{\partial
y^2} +\frac{1}{4}\left(r^2+\lambda^2 y^2\right) \nonumber \\ &+&
{L^2-1\over r^2} + 8\sqrt 2 \pi n\left|\frac
{\varphi({r,y};t)}{r}\right|^2  \biggr]\varphi({ r,y};t)=0,
\end{eqnarray}
where the length, time, and  wave function are expressed in units of 
$\sqrt {\hbar/(2m\omega)}$, $\omega ^{-1}$, and $ [r\sqrt{{l^3}/{\sqrt
8}}]^{-1}$, respectively. Here radial and axial trap frequencies are 
$\omega$ and $\lambda\omega$, respectively, $m$ is the atomic mass, 
 $l\equiv \sqrt {\hbar/(m\omega)}$ is the harmonic oscillator length, and 
$ n =   N a /l$ the nonlinearity with $a$ the interatomic scattering
length. 
For solitonic states $n$ is negative.
In terms of the
one-dimensional probability
$P(y,t)$ defined by
\begin{equation}\label{avi}
P(y,t) = 2\pi \int_0 ^\infty
dr |\varphi(r,y,t)|^2/r , 
\end{equation}
the normalization of the wave function
is given by $\int_{-\infty}^\infty dy P(y,t) = 1$

We solve the GP equation (\ref{d1}) numerically using 
split-step time-iteration method using the Crank-Nicholson discretization
scheme described recently \cite{11}. The details of the numerical scheme
for this problem can be found in  \cite{njp}. 


{\underline{Results:}} For bright solitons the nonlinearity $n$ is
negative. Under the
conditions $n<0$  and   $\lambda = 0$, a soliton-type BEC state
can be generated only for $n$ greater than a critical value
($n_{\mbox{cr}}$):  $n_{\mbox{cr}}< n<0$. For $n<n_{\mbox{cr}}$,
the system becomes too attractive and collapses and no stable soliton
could be generated.  
The actual value of $n_{\mbox{cr}}$ is a function
of the trap parameter $\lambda$. For the spherically symmetric case
$\lambda = 1$, and  $n_{\mbox{cr}}= -0.575$ \cite{8,9}.  

Numerically solving the GP equation (\ref{d1}) for $\lambda=0$, we find
that  the
critical
$n$ for collapse of a single soliton is
$n_{\mbox{cr}}= -0.67$ for $L=0$ and
 $n_{\mbox{cr}}=  -2.10$ for $L=1$. For  $L=0$ $n_{\mbox{cr}}= -0.67$
is in close agreement with 
with $n_{\mbox{cr}}= -0.676$  obtained by other workers
\cite{4a0,crit}. For $L=1$,  $n_{\mbox{cr}}=  -2.10$ should be contrasted 
with $n_{\mbox{cr}}=  -2.20$ obrained by Salasnich \cite{bvs}.
For $L=0$,
$\omega = 2\pi\times 800$ Hz
and final scattering length $-3a_0$
as in the experiment of Strecker {\it et al.} \cite{3},  $n_{\mbox{cr}}=
-0.67$ corresponds to about 6000 $^7$Li atoms. One can have
proportionately about
three times more atoms in the $L=1$ state.

\begin{figure}
 
\begin{center}

\includegraphics[width=0.49\linewidth]{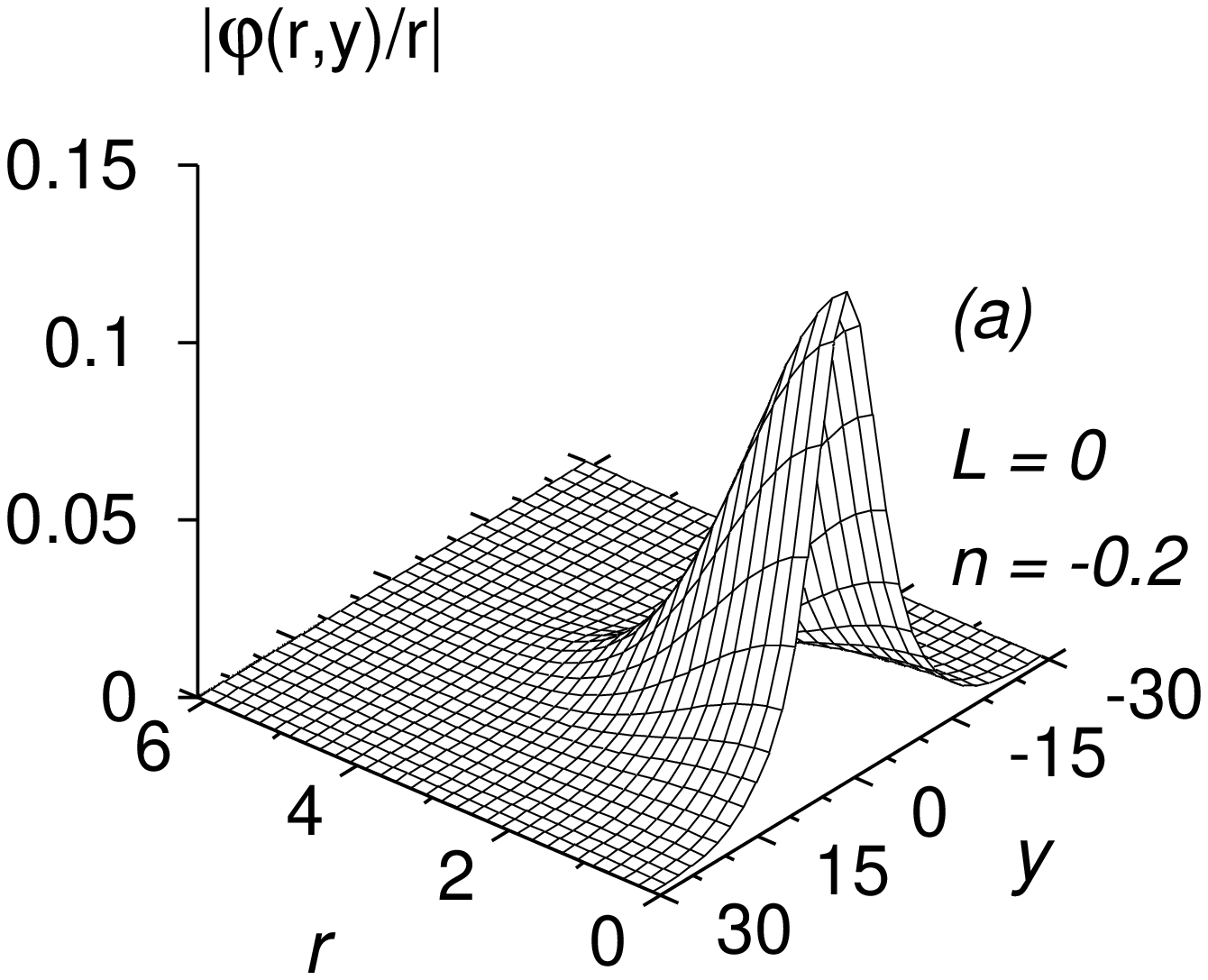}
\includegraphics[width=0.49\linewidth]{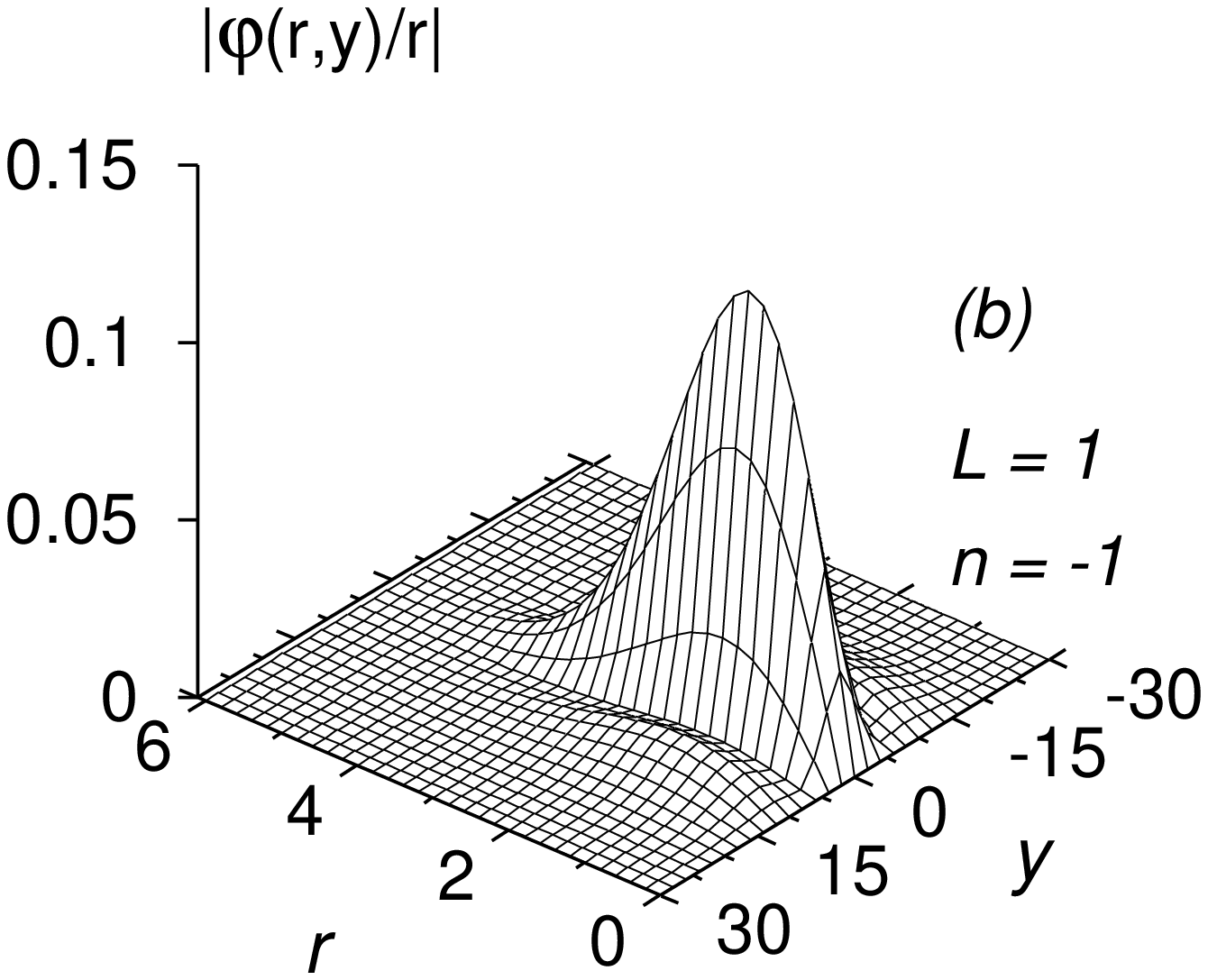}
\end{center}
 
\caption{Three-dimensional  wave function  $|\varphi
(r,y)/r|$
vs. $r$ and $y$
for a single soliton  with $\lambda
= 0$, and   (a) $L=0$, $n=-0.2$, and (b) $L=1$, $n=-1$. }
\end{figure}

A $L=0$ soliton with $n=-0.2$ is illustrated in Fig. 1 (a) where we plot
the three-dimensional wave function $|\varphi (r,y)/r|$ vs. $r$ and $y$.
For  $L=1$ we calculated the soliton for
$n=-1$
and plot  $|\varphi (r,y)/r|$ in Fig.
1 (b).  Because of the radial trap the soliton remains confined in the
radial direction $r$, although free to move in the axial $y$
direction. The nature of the two wave functions are different. For $L=0$,
the condensate has maximum density for $r=0$. For $L=1$, because of
rotation a vortex has been generated along the axial direction
corresponding to a zero density for  $r=0$.

Next we consider the interaction between two solitons with a
phase difference $\delta$ given by the following superposition of two
solitons $\bar\varphi$ at $\pm y_0$ at time $t=0$
$\varphi(r,y)=|\bar \varphi(r,y+y_0)|+ e^{i\delta}
|\bar \varphi(r,y-y_0)|.$
The time evolution of these two  solitons is found using the
solution of (\ref{d1}) for different $\delta$.
In the present simulation we consider two $L=1$
equal vortex solitons each of $n = -0.4$ for $\lambda =0$ initially at
positions $y_0=\pm 15$ 
and observe them for an interval of time $t=400$. We also consider the
evolution of 
two
$L=0$ solitons each of $n = -0.2$ for $y_0=\pm 15$.
The solitons    interact by exchanging 
particles and after an interval of time two unequal solitons are generated
from two equal solitons. The evolution of the two solitons in the $L=0$
case for $\delta = \pi/2$ is shown in Fig. 2. Similar evolution for the
$L=1$ case is reported in \cite{njp}.


\begin{figure}
 
\begin{center}
\includegraphics[width=0.47\linewidth]{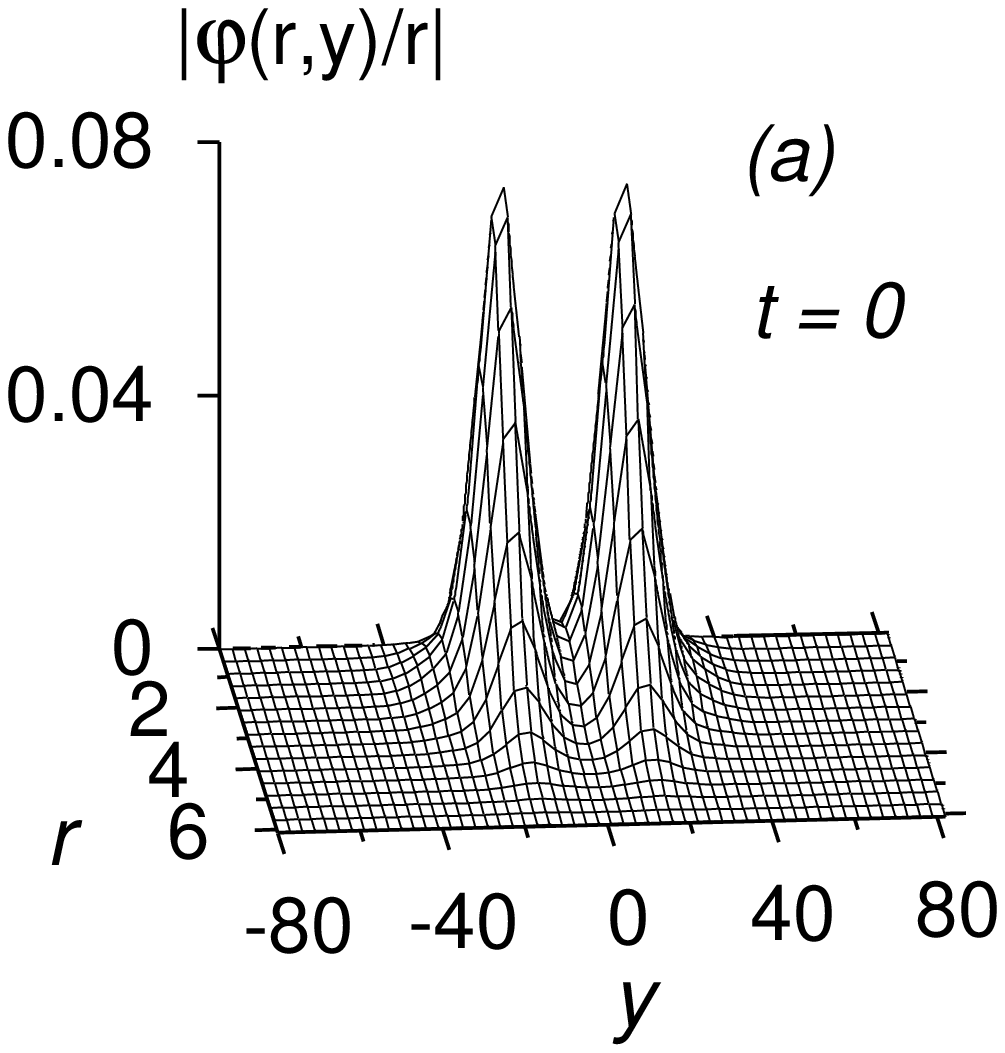}
\includegraphics[width=0.47\linewidth]{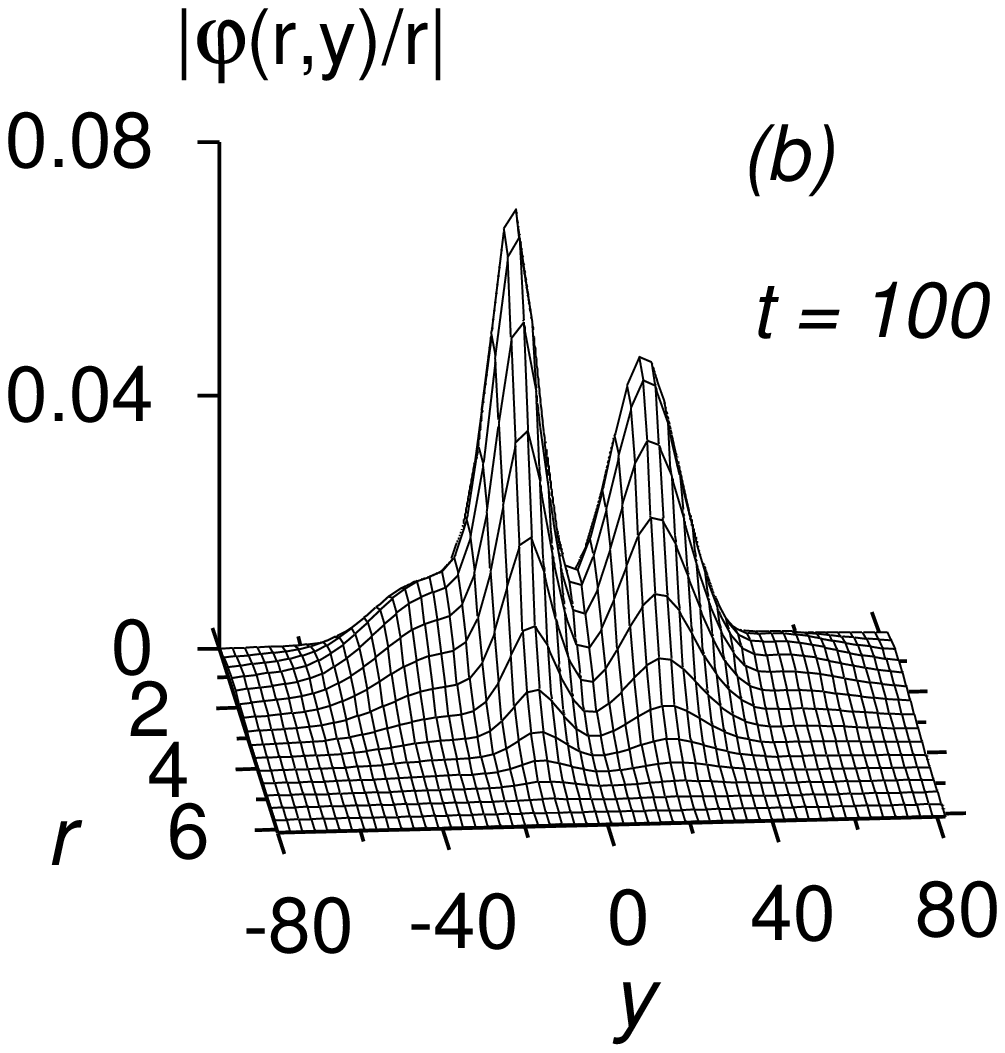}
\includegraphics[width=0.47\linewidth]{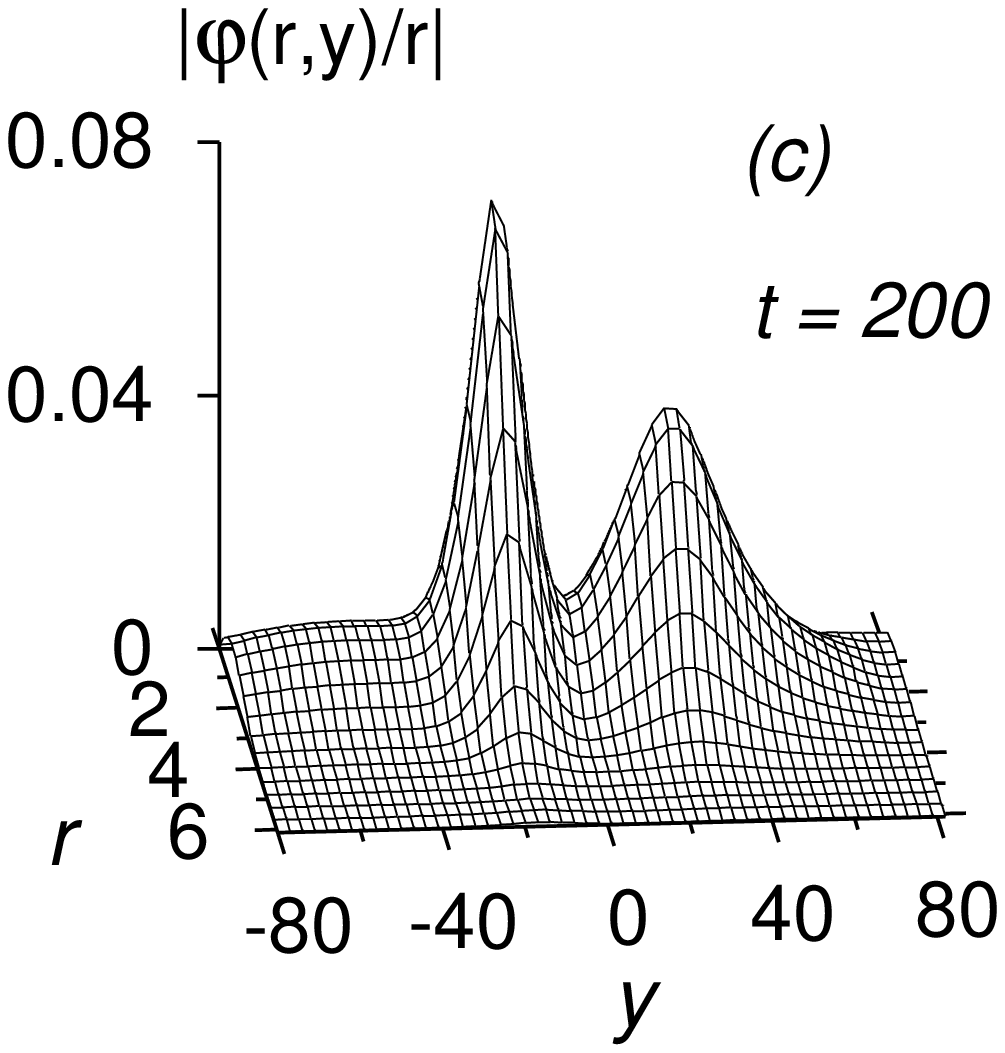}
\includegraphics[width=0.47\linewidth]{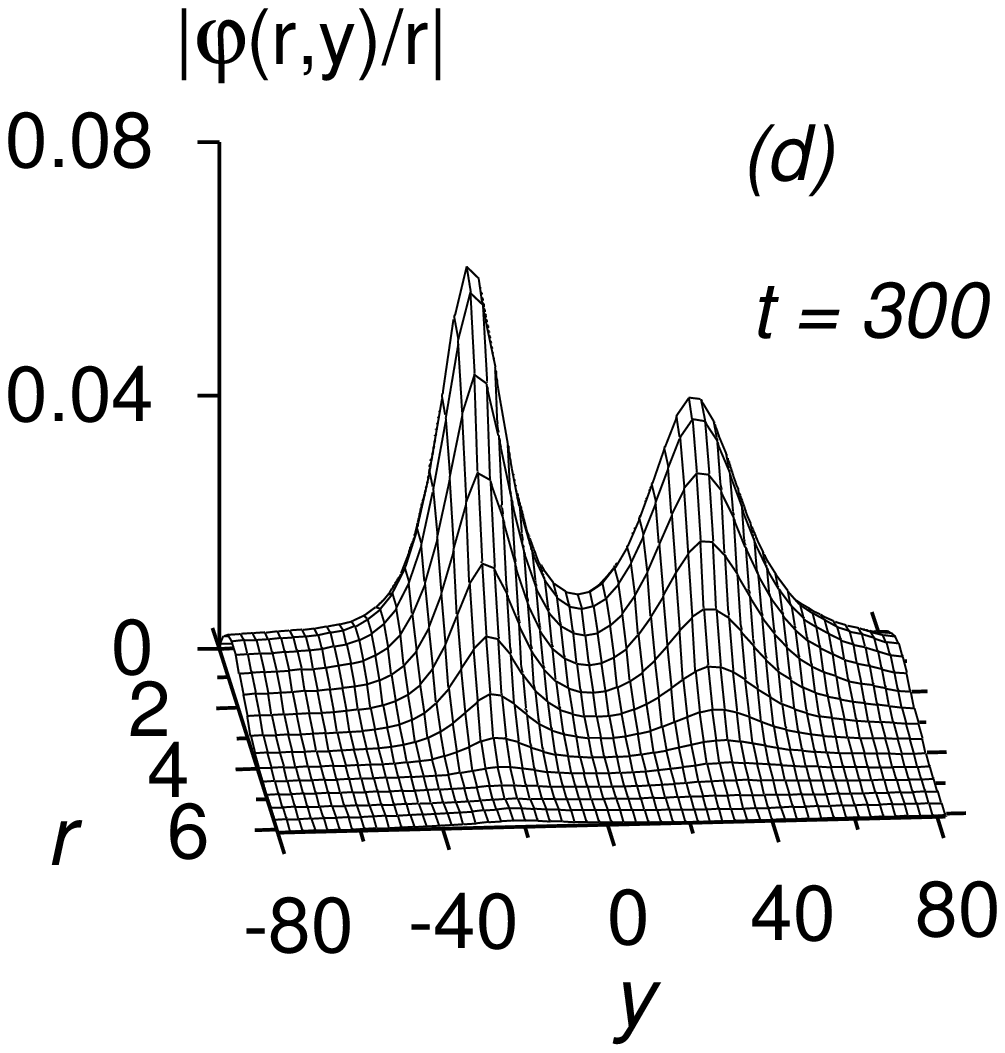}
\end{center}
 
\caption{The wave function $|\varphi(r,y)/r|$ vs. $r$ and $y$ of two $L=0$
solitons with $n=-0.2$ each and $\delta =\pi/2$ at times (a) $t=0$,
(b) 100, (c) 200
and (d) 300.}
\end{figure}

\begin{figure}
 
\begin{center}
\includegraphics[width=0.47\linewidth]{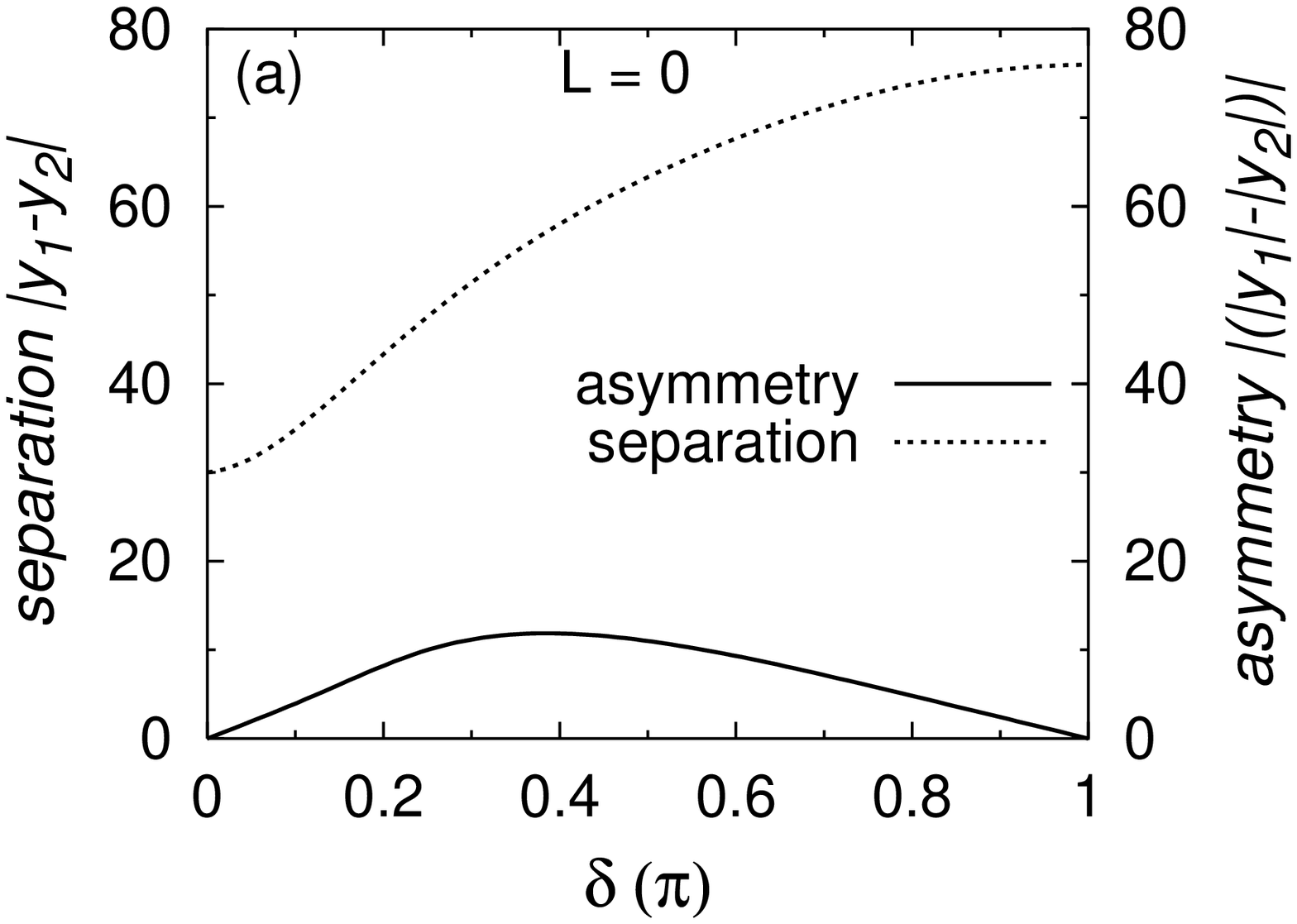}
\hskip 0.5cm
\includegraphics[width=0.47\linewidth]{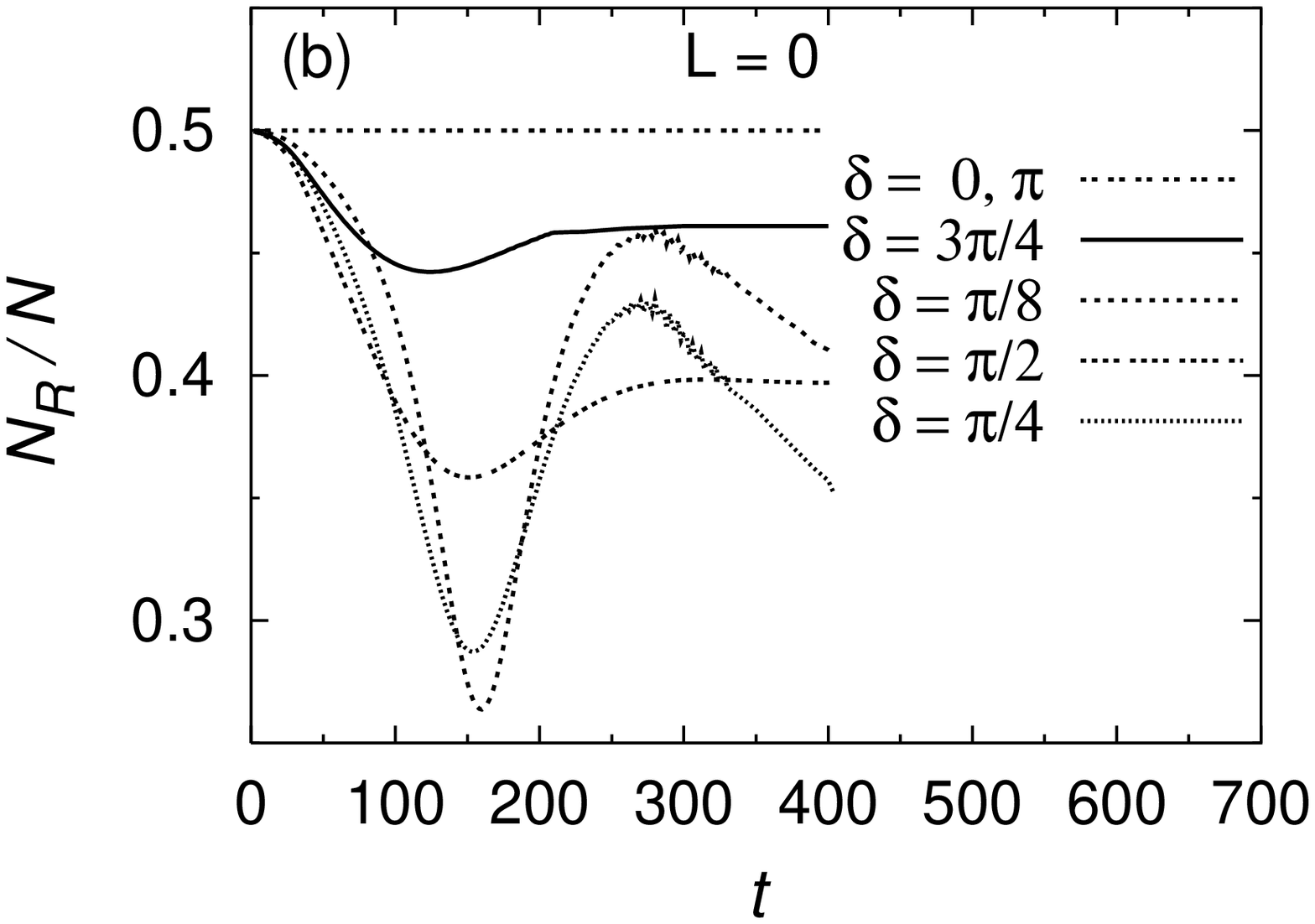}
\includegraphics[width=0.47\linewidth]{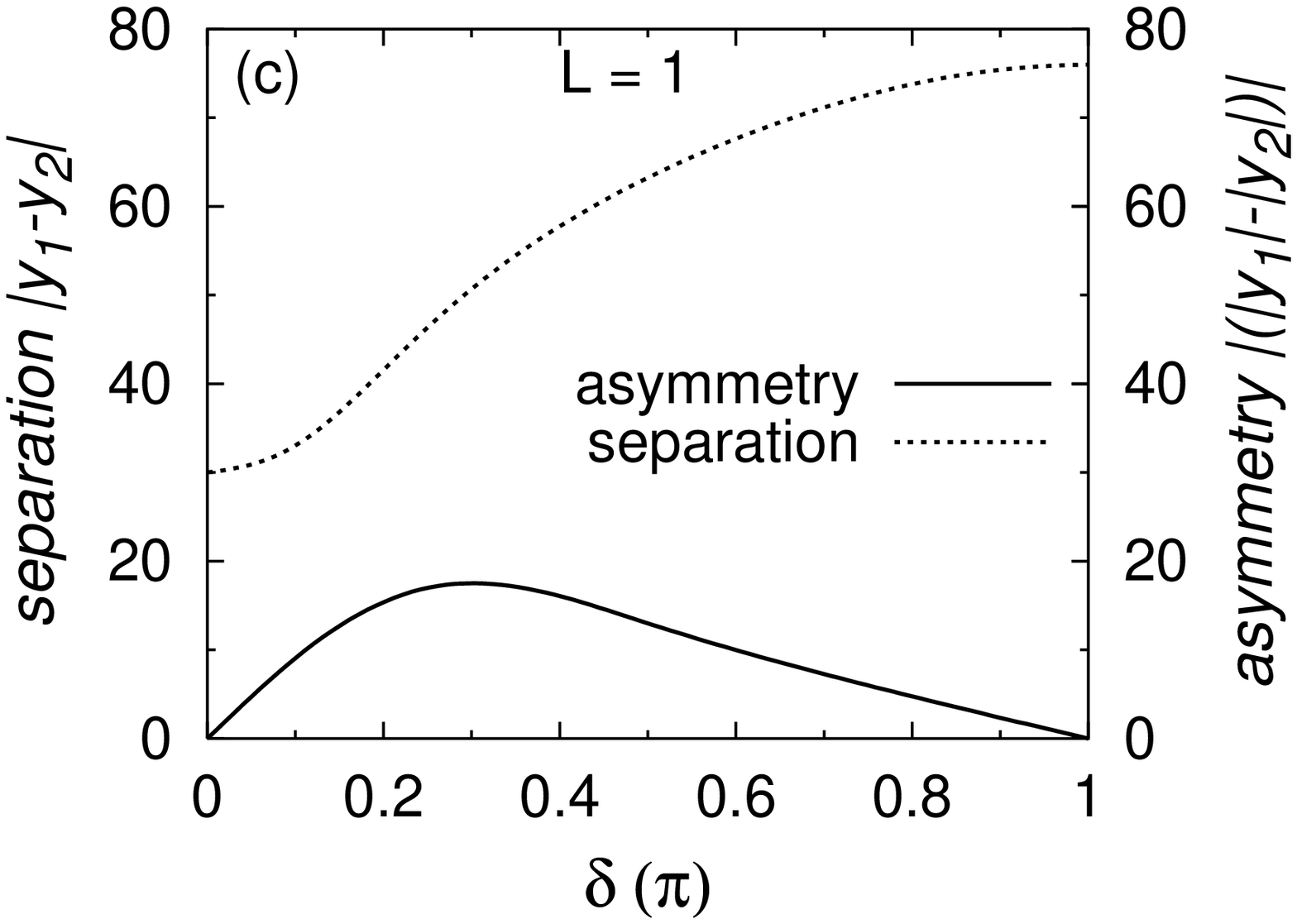}
\hskip 0.5cm
\includegraphics[width=0.47\linewidth]{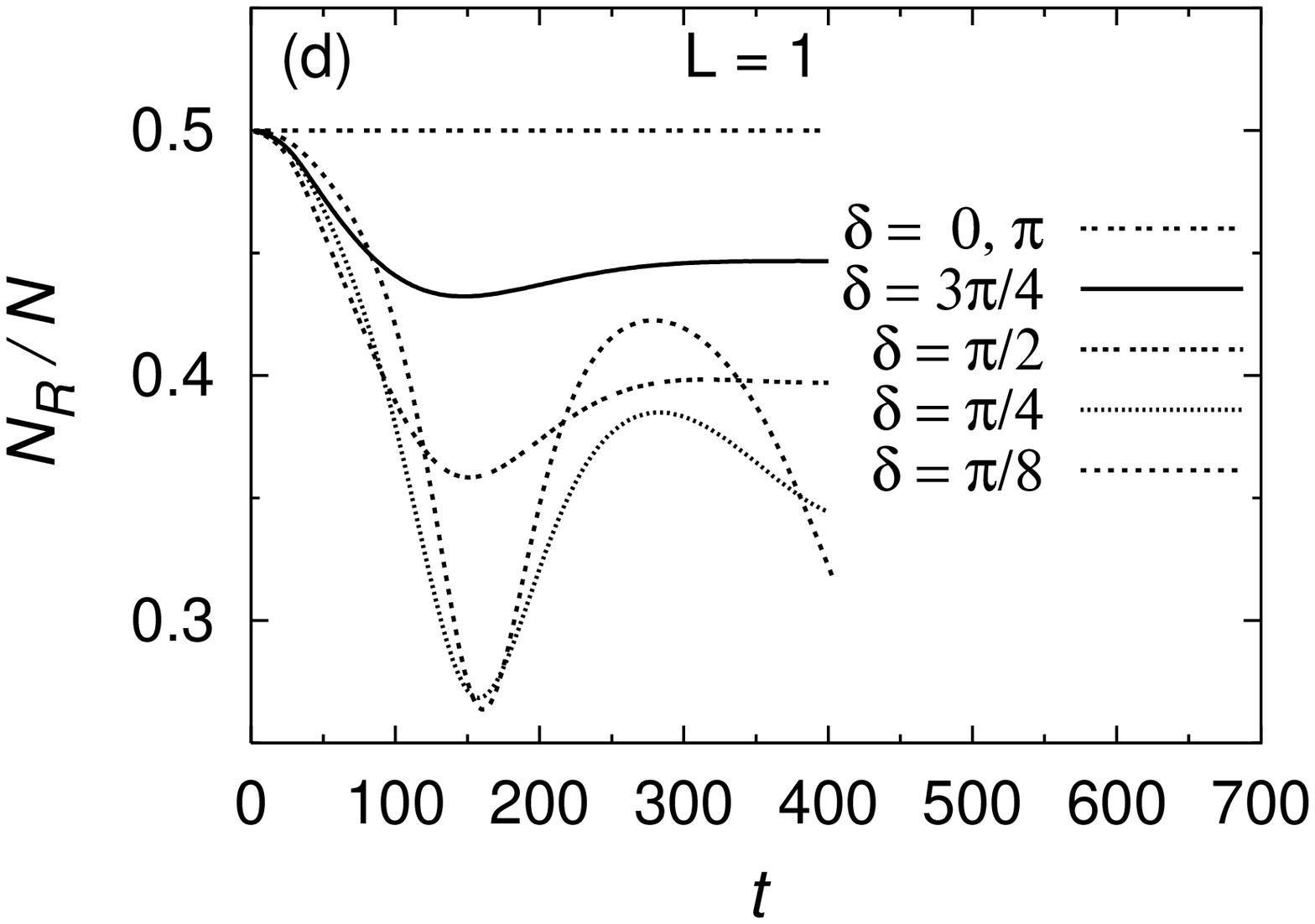}
\end{center}
 
\caption{(a) Final asymmetry and separation in the position of the two
solitons of $L=0$ and $n=-0.2$ each at $t=400$ vs. phase
$\delta$. (b) Ratio $N_R/N$ vs. $t$ for different phase $\delta$ for two
$L=0$ and  $n=-0.2$ solitons initially at $\pm 15$. $N_R$ is the number
of atoms in the right soliton and $N$ the total number of atoms.
(c) Same as (a) for two solitons  of $L=1$ and $n=-0.4$. (d)  Same as (b)
for two solitons  of $L=1$ and $n=-0.4$.}
\end{figure}

From Fig. 2 we find that at $t=0$ the two solitons are equal  and
symmetrically located.  However, this symmetry is broken for $t>0$.
The asymmetry and separation in the final position of the solitons $y_1$
and $y_2$ are best studied via $ |(|y_1|-|y_2|)| $ and $|y_1-y_2|$,
respectively,
at time $t=400$ for different phase $\delta$ between the solitons and in
Fig. 3 (a) we plot the same for different $\delta$ for the $L=0$ case. We
find that the
asymmetry is zero for $\delta=\pi$ and 0 and is largest for a $\delta$ in
between. However the separation increases monotonically as $\delta$
increases from 0 to $\pi$. Hence the interaction is repulsive for almost
all  $\delta$     except for  $\delta\approx 0$.      
Closely associated with the asymmetry and
separation in the final position of the solitons is the number of
exchanged atoms between the two solitons, which demonstrates the change in
the sizes of the solitons. Actually, the smaller soliton travels faster
and the larger one travels slower. This results in the asymmetry in the
final positions. The change in the sizes of the solitons for $L=0$ is
demonstrated in the plot of $N_R/N$ vs. $t$
for different $\delta$ in Fig. 3 (b), where $N_R$ is the number of atoms
in the right soliton and $N$ the total number of atoms in the two
solitons.  In Figs. 3 (c) and (d) we plot the same for $L=1$.
The variation of $N_R/N$ is qualitatively  similar for $L=0$ and 1 in
Figs. 3 and also 
to that found in
\cite{6a01} for $L=0$. However, there are quantitative differences,
specially at large times.   
In the present simulation we find that,
for the change $\delta \to -\delta$, $N_R/N \to N_L/N$, where $N_L\equiv
(N-N_R)$ is the number of atoms in the left soliton.

If the phase difference $\delta$ between two
neighboring solitons is not close to zero, they experience overall
repulsion  and
stay apart. However, for $\delta$ close to zero they interact attractively
and often a soliton could be lost as observed in the experiment of
Strecker {\it et al.} \cite{3}.  
Throughout this investigation in the interaction of two equal solitons we
assumed that the nonlinearity $|n|$ for each is less than
$|n_{\mbox{cr}}|/2$, so that a stable solitonic condensate with total $|n|
<|n_{\mbox{cr}}|$ exists when the two  coalesce. However, if two solitons
each with  $|n| > |n_{\mbox{cr}}|/2$ encounter for $\delta =0$, the system
is expected to  coalesce,
collapse and emit atoms via three-body recombination. It is
possible that in this case only a smaller single soliton survives.  This
might
also explain some  missing soliton(s) in experiment.

\section{Conclusion}

We emphasize  the possibility of creating and studying bright vortex
solitons of an attractive  BEC in laboratory under radial trapping. We
determine the condition of critical stability of bright solitons.  
Employing a numerical solution of the GP equation with axial
symmetry, we have performed a realistic mean-field study of the
interaction
among two bright solitons in a train and find the overall interaction to
be repulsive except for phase $\delta $ between neighbors close to 0.
There
is an inelastic exchange of atoms between two solitons resulting in a
change of size and shape.  Except in the $\delta \approx 0$ case, the
solitons in a train stay apart and never cross each other as observed in
the experiment by Strecker {\it et al.} \cite{3}.  For $\delta \approx 0$
a single soliton can often disappear as a result of the attractive 
interaction among solitons, as observed experimentally by Strecker {\it et
al.}. 
The $L=1$ vortex solitons can accommodate a larger number of 
atoms and the present study may motivate future experiments with them.  

\begin{acknowledge}
Work partially supported by the CNPq of Brazil.
\end{acknowledge}

\end{document}